\documentclass[11pt] {article}
\usepackage{amsmath} 

\newcommand{\mysection}[1]{\setcounter{equation}{0}\section{#1}}

\def\cc#1{\kern .7em\hfill #1 \hfill\kern .7em}  

\def\CC{\hbox{\it l\hskip -5.5pt C\/}}
\def\RR{\hbox{\it I\hskip -2.pt R }}
\def\MC{\hbox{\it I\hskip -2.pt M \hskip -7 pt I \hskip - 3.pt \CC}_n}
\def\MMC{\hbox{\it I\hskip -2.pt M \hskip -7 pt I \hskip - 3.pt \CC}}
\newcommand{\nc}{\newcommand}
\nc{\beqa}{\begin{eqnarray}}
\nc{\eeqa}{\end{eqnarray}}
\def\agt{
\mathrel{\raise.3ex\hbox{$>$}\mkern-14mu\lower0.6ex\hbox{$\sim$}}
}
\def\alt{
\mathrel{\raise.3ex\hbox{$<$}\mkern-14mu\lower0.6ex\hbox{$\sim$}}
}

\begin {document}
\bibliographystyle{unsrt}    

\hspace{13cm}PM$-99$/$33$

\hspace{13cm}KIAS-P99082

\Large

\vspace{2cm}
{\centerline{\bf Extended Complex Trigonometry in Relation to}}
\vspace{3mm}

{\centerline{\bf  Integrable $2D-$Quantum Field Theories and Duality }}
\vspace{3mm}

\large

\vspace{1cm}

\centerline{P.~Baseilhac$^{1,2,3}$, S.~Galice$^{1}$, P.~Grang\'e$^{1}$}

\vspace{3mm} 

\centerline{and}

\vspace{3mm}

\centerline{M.~Rausch de Traubenberg$^{1,4}$}

\small\normalsize
\vspace{0.5cm}

\centerline { $^{1}$Laboratoire de Physique Math\'ematique, Universit\'e 
Montpellier II}

\vspace{2mm}

\centerline {Place E.~Bataillon, 34095 Montpellier, France}

\vspace{2mm}

\small\normalsize
\vspace{2mm}

\centerline { $^{2}$School of Physics, Korea Institute of Advanced Study }

\vspace{2mm}

\centerline {Seoul, 130-012, Korea}

\vspace{2mm}

\centerline { $^{3}$Department of Mathematics, University of York }

\vspace{2mm}
\centerline {Heslington, York YO105DD, United Kingdom}

\vspace{2mm}

\small\normalsize

\vspace{2mm}

\centerline { $^{4}$ Laboratoire de Physique Th\'eorique}

\vspace{2mm}

\centerline {3 rue de l'Universit\'e, 67084 Strasbourg, France}

\vspace{1cm}

\begin{abstract}
Multicomplex numbers of order $n$ have an associated trigonometry
(multisine functions with $n-1$ parameters) leading to a natural 
extension of the sine-Gordon model. The parameters are constrained 
from the requirement of local current conservation. In two dimensions 
for  $n < 6$  known integrable models (deformed Toda and non-linear 
sigma, pure affine Toda...) with dual counterparts are obtained in 
this way from the  multicomplex space $\MC$ itself and from the 
natural embedding $\MC \subset \MMC_m, n < m$. For $ n \ge 6$ a 
generic constraint on the space of parametersis obtained from 
current conservation at first order in the interaction Lagragien.
\end{abstract}

\vspace{1mm} 

\newpage
\pagestyle{plain} 

\setcounter{page}{2}
\mysection{Introduction}\label{Introduction}
In the description of strongly interacting systems most of the
relevant physical quantities such as spectra, correlation functions ...,
cannot be obtained by perturbative methods and many techniques have been 
investigated to go beyond the regime of weak coupling. In gauge theories the
electric-magnetic duality has been pointed out \cite{mool,sewit} as
a possibility to relate weak and strong coupling situations. It is by now
common to speak of duality only in this sense and a well known example in
two dimensions is the  link between the sine-Gordon and massive Thirring models
 \cite{Col}. In a more recent past, in order to understand further
the mechanisms of duality in mass spectra and $S-$matrices, the
attention focussed on Toda systems. They are examples of integrable though
deformed conformal field theories with affine Lie algebras symmetries
showing exact duality \cite{ArFaVa}.  
An ever growing number of such deformed Toda models  is currently proposed 
\cite{Fateev1}.
The question naturally arises of the existence of an underlying 
common structure suitable for a general analysis. The issue was studied in a
 recent publication
\cite{muli4}.
 This structure
comes from  the introduction
of Clifford algebras of polynomial and generalized Clifford algebras 
\cite{line}. 
These algebras are associated to the linearization of polynomials of degree 
$n$, 
in the same way the usual Clifford algebra is associated to
Dirac's linearization of the laplacian operator. 
These Clifford families of order $n$ contain a natural extension of  complex 
numbers
and trigonometric functions, called respectively
multicomplex numbers and  multisine-functions \cite{mc1}. 
A generalization of the sine-Gordon model in the multicomplex space
$\MC$ was proposed in \cite{muli4}.  A first investigation of  
integrability properties in two space-time dimensions was also given. In this 
procedure a 
necessary step is the  construction and analysis of conserved quantities.
Because of their evident link with integrable cases only low $n$ cases 
($n \le 4$) were then studied. 
Here, new interesting properties of the multisine functions are exhibited 
which lead to a more general analysis.
On the one hand a natural embedding of the multicomplex
space exists which leads to ``non-minimal'' multisine-functions as 
opposed to the ``minimal'' ones considered in \cite{muli4}.
New multisine-Gordon models emerge which can be either identified to
known integrable cases with dual counterparts or are  truly new.
On the other hand the multisine-functions can be written in a generic
way which gives access to hidden symmetries and permits the 
disentanglement of  the elaborated algebra associated with non-local conserved 
charges \cite{pascal}.

The purpose of this letter is to generalize the analysis of \cite{muli4}.
We shall leave apart most technicalities, but focus on some aspects in 
relation to
duality and current conservation for arbitrary dimension of the multicomplex
space $\MC$.

To start, a brief summary  of the definition of the  multisine-functions is
given. They naturally contain $n-1$ parameters and their properties are given 
in relation
to the different underlying possible structures of the MC-numbers. 
The conservation of {\it local} currents implies relations 
between the parameters. For real parameters specific solutions
are identified to known integrable QFT's and their dual representations are 
exhibited. Some cases with imaginary parameters have original dual properties. 
An example 
is discussed for $n=4$. Perspectives and conclusions are then drawn.

\mysection{Multicomplex numbers and multisine functions}

Multicomplex (MC) numbers \cite{mc1} have been introduced in the past in 
relation to generalized
Clifford algebra sustained by linearization of polynomials of degree higher 
than two \cite{line}. 
The starting point is the introduction
of a generator $e$, such that $e^n=-1$, sustaining the space
of multicomplex numbers of order $n$, $\MC$. In keeping with the case $n=2$ of
 usual complex numbers and
their trigonometric functions, an associated extended trigonometry follows. It
 is characterized by specific
``angular'' functions dubbed multisine (mus). These $\mathrm{mus}$-functions 
depend on 
$\phi_a, a=0 \cdots, E(n/2)-1$ compact quantities, and
$\varphi_a$ $a=0,\cdots,n- E(n/2)-1$ non-compact ones ($E(a)$ is the integer
 part of $a$). A collection of useful relations 
\cite{mc1} exists
between the mus-functions: additions, derivatives etc.... \\
At variance with  the quadratic case, where the sine and cosine functions are 
unique,
when $n>2$, different type of functions may be obtained. However when obtained
 as for the 
case $n=2$ from the exponentiation of MC-numbers \cite{muli4,mc1,pascal,galice}
, all these mus-functions can be written under a common form.
They are connected to the existence of
natural embeddings of the multicomplex space $\MC \subset \MMC_m, n < m$,
with any $n$ for $m$  odd and  even $n$ when $m$ even. Indeed, starting from
$e_m, ((e_m)^m= -1)$, the generator of $\MMC_m$, it is possible to define
several
generators of $\MC, e_{(n|m)}$, satisfying  $(e_{(n|m)})^n = -1$ \cite{galice}.
But now, if the algebra $\MMC_{(n|m)}$ is considered as a sub-algebra of
$\MMC_m$, the associated mus-functions satisfy  an $m-$th order relation 
instead
of an $n-$th order one.
 In orther word, these
mus-functions of order $n$ are related to those of higher order $m$.
 They are given as \\
\beqa
\label{mus0n4}
\left\{
\begin{array}{lll}
\mathrm{mus}_{k}\left(\{\phi_a\},\{\varphi_a\}\right) = &{\frac 2 n}  
\sum \limits_{a=0}^{n/2-1}
\cos( \phi_a - (2a+1) {k \pi \over n}) \exp( \varphi_a)
&\mathrm{for~} n \mathrm{~even} \\
\mathrm{mus}_{k}\left(\{\phi_a\},\{\varphi_a\}\right) = &{\frac 2 n}  
\sum \limits_{a=0}^{(n-1)/2-1}
\cos( \phi_a - (2a+1) {k \pi \over n}) \exp( \varphi_a) +& \\
&{ (-1)^k \over n} \exp{\Big( \varphi_{(n-1)/2}\Big)}
&\mathrm{for~} n \mathrm{~odd}. \\
\end{array}
\right.  
\eeqa
These mus-functions differ from each other only through the constraint
 necessary to ensure
 unimodular MC-numbers ({\it cf }  $ \|z\|=1$ for $n=2$) and which connects 
the non-compact variables.
The $\varphi_a$'s must verify
\begin{equation}
\label{constr_m}
\left\{
\begin{array}{ll}
\sum \limits_{a=0}^{n/2-1} m_a \varphi_a=0&{\mathrm{for~}}n{\mathrm{~even}} \\
 \sum \limits_{a=0}^{(n-1)/2-1} 2 m_a \varphi_a + m_{(n-1)/2}
 \varphi_{(n-1)/2} = 0
&{\mathrm{for~}}n{\mathrm{~odd}} \\
\end{array}
\right.
\end{equation}
with $2\sum \limits_{a=0}^{n/2-1} m_a=m$ (for even $n$) and by
$2\sum \limits_{a=0}^{(n-1)/2-1}  m_a+ m_{(n-1)/2}=m$  
(for odd $n$)\cite{galice}. 
 The cases with $m_a=1, a=0,\cdots, E({ n-1 \over 2})$
correspond to particular mus-functions, called `minimal'' in the sequel.

\mysection{Duality in multisine-Gordon models}

Without loss of generality the multisine-Gordon model $MSG(n|m)$ for $d=2$  
is defined   \cite{muli4,pascal,galice}  in terms of the function 
$\mathrm{mus}_0$. 
The action writes 
\begin{eqnarray}
{\cal{A}}^{(n|m)} = \int\!\!d^2x \Big[{{1} \over {16\pi}} \big[ 
\sum \limits_{a=0}^{E(n/2)-1} 
\!\!\!\left(\partial_{\mu}\phi_a\right)^2  
+\sum \limits_{a=0}^{n-E(n/2)-2}
\!\!\!\left(\partial_{\mu}\varphi_a\right)^2 \big] 
 -\mu\mathrm{mus}_0(\{\alpha_a\phi_a\},\{\beta_a\varphi_a\})\Big],
\label{action}
\end{eqnarray}
with $E(n)$ the integer part of $n$, $\mu$  real, ($\alpha_a$,$\beta_a$) 
complex parameters and $(n|m)$ stands for the ``embedding'' 
 considered.
This model is denoted as $MSG_{(n|m)}(\{\alpha_a\},\{\beta_a\})$. 

Following the approach of A. Zamolodchikov \cite{Zamo1},
quantum integrability in $2d$ is related to the conservation of a non-trivial 
current (spin higher than two).
The analyzis can be carried out from massless perturbation theory in which the
 whole exponential terms are treated as interaction terms, 
with tadpoles divergences removed by normal ordering\,\footnote{The whole set 
of parameters is assumed to satisfy 
the  renormalizibility condition {\it i.e.} no other vertex operator terms are
 generated under RG 
group flow;
then conservation to  first order in the interaction term
is sufficient to ensure integrability.}.
 A conjecture is made for local holomorphic currents in terms of a
sum of all possible operators $(\partial^p \Phi_a)^{n_1} 
(\partial^q \Phi_b)^{n_2}$ 
($\Phi=(\phi,\varphi)$) of a given overall spin
$s$ compatible with the symmetries of the lagrangian and with specific
 coefficients
$X_{a,p,b,q}^{n_1,n_2}$. An operator product expansion (OPE) is then performed 
for the action of the
current on the operator vertices. In conformal perturbation theory (CPT) the
 conservation of the current  
to  first order in the interaction term
requires the necessary vanishing of all
terms which cannot be expressed as exact derivatives. 
A set of linear equations results for 
the  coefficients $X_{a,p,b,q}^{n_1,n_2}$. For the multisine-Gordon models 
this system has  three types of solutions leading to constraints on the 
parameters
$\alpha_a$ and $\beta_a$.
Solutions corresponding to particular constant values of $\alpha_a$ and 
$\beta_a$ are exotic
and will not  be analyzed.  The second type is characterized  by a mixing of 
real and complex
 values
of the parameters satisfying a quadratic equation. Some of these cases 
correspond to Affine Toda
Field Theories
(ATFT) discussed in the next section. The last type is given by {\it real} 
values
of the parameters subject to quadratic constraints.

For these real values the solvability conditions correspond to an hypersurface
 in the parameter
space $\{\alpha_a,\beta_a\}$ which turns out to have in general a simple 
geometrical interpretation. As we
shall see the embedding property of the multicomplex space permits now to 
take into account 
 known perturbed CFT
and leads to new models which could not be tackled in \cite{muli4}. 
This family of models are not only specified by $n$ and $m_a, a=0,\cdots
 E({n-1 \over 2})$,
but also by the relation(s) among the parameters $\alpha_a$ and $\beta_a$.

The type  of solutions given by {\it real}
 values of the parameters subject to quadratic constraints and showing 
interesting duality properties is now discussed. 

For $n=3$  the current of  spin $4$ is conserved provided the parameters obey
  the
constraints
\begin{eqnarray}
\label{p3m}
\begin{array}{lll}
(i)& \alpha_0^2-\beta_0^2 = 1/2,& m_0/m_1=1,1/2 \\
(ii)&{m_0 \over m_1}\big[\alpha_0^2-\beta_0^2\big]-{m_0 \over m_1}\big[2 {m_0
 \over m_1}
\beta^2_0\big]-\big[\frac{1}{2}+{m_0 \over m_1}\big]=0,& m_0,m_1 \ne 0.
\end{array}
\end{eqnarray}
The case $(i)$ with $m_0/m_1=1$, (minimal representation)  was already 
envisaged  in \cite{muli4}.
Its integrability and dual properties were derived in \cite{Fateev1}, but 
outside the present general context.
The case  with $m_0/m_1=1/2$ 
corresponds to the embedding
$\hbox{\it I\hskip -2.pt M \hskip -7 pt I \hskip - 3.pt \CC}_3 \subset 
\hbox{\it I\hskip -2.pt M \hskip -7 pt I \hskip - 3.pt \CC}_4$. Its duality
property will be discussed below.

For $n=4$ a spin $4$ current is conserved if
\begin{eqnarray}
\begin{array}{lll}
(i)&\left\{
\begin{array}{l}
\alpha_0^2-\beta_0^2=1/2 \\
\alpha_0=\alpha_1
\end{array}
\right.&m_0/m_1=1\\
& \\
(ii)& \left\{
\begin{array}{l}
\alpha_0^2-\beta_0^2=1/2 \\
\alpha_1^2- {1 \over 4 \beta_0^2} = 1/2 
\end{array}
\right.&m_0/m_1=1/(2 \beta_0^2) \\ 
& \\
(iii) & \frac{m_0}{m_1}\big[\alpha_{0}^{2}-\beta_{0}^{2}\big]
+\big[\alpha_{1}^{2}-\frac{m^2_0}{m^2_1}\beta_{0}^{2}\big]
-\big[1+{m_0 \over m_1}\big]=0,& m_0,m_1 \ne 0.   
\end{array}
\label{cont4m}
\end{eqnarray}
The first solution has properties similar to the earlier case $(i)$ 
\cite{Fateev1}. 
Solution $(ii)$ has a self-dual representation 
for which the general discussion of the next section is relevant. 

Finally, for $n=5$  only  the generic solutions  
and  one already studied in \cite{Fateev1} (of type $(i)$-Eq.(\ref{p3m}) or 
$(i)$-Eq.(\ref{cont4m})) will be discussed. 
The conditions are respectively
\begin{equation}
\begin{array}{lll}
(i)& \alpha_0^2 - \beta_0^2=1/2,~ \alpha_0=\alpha_1, 
\beta_0=-\beta_1,& m_0/m_2=m_1/m_2=1/2 \\
 \\
(ii)&
\frac{m_0}{m_2}\big[\alpha_{0}^{2}-\beta_{0}^{2}\big]
+\frac{m_1}{m_2}\big[\alpha_{1}^{2}-\beta_{1}^{2}\big] \\
&-\frac{m_0}{m_2}\big[2\frac{m_0}{m_2}\beta_{0}^{2}\big]
-\frac{m_1}{m_2}\big[2\frac{m_1}{m_2}\beta_{1}^{2}\big]
-\big[ \frac{1}{2}+ \frac{m_0}{m_2}+\frac{m_1}{m_2}\big]=0,& m_0,m_1,m_2 
\ne 0.  
\end{array}
\label{cont5m}
\end{equation}
The first one corresponds to 
$\hbox{\it I\hskip -2.pt M \hskip -7 pt I \hskip - 3.pt \CC}_5 \subset 
 \hbox{\it I\hskip -2.pt M \hskip -7 pt I \hskip - 3.pt \CC}_6$.  This 
QFT possesses a dual
representation discussed below.

With  these constraints the current of spin $4$ is conserved 
to  first order  in CPT. This is enough to ensure integrability under certain 
conditions ({\it cf } footnote $1$).
The models listed in the table below have this property.

The families of QFT's
associated to the conditions $(i)$ in (\ref{p3m}) and $(i)$ in (\ref{cont4m})
 have a simple interpretation. In these two
cases, in the action (\ref{action}) 
 one may always consider the terms in $\exp(\beta_0 \varphi_0) \cos(\alpha_0
 \phi_0)$
in $\mathrm{mus}_{0}\left(\Phi\right)$ together with the kinetic contributions
 and with the condition 
$\alpha_0^2-\beta_0^2 = 1/2$ as an initial conformal field theory (CFT) 
deformed by the remaining part
of the  $\mathrm{mus}_{0}\left(\Phi\right)$.\\

For  an appropriate comparison with known results a new normalization is now 
retained such that 
($\alpha_a \rightarrow \alpha_a/\sqrt{8\pi}$, $\beta_a \rightarrow 
\beta_a/\sqrt{8\pi}$). For small 
parameters,
 it is possible to fermionize the QFT's using the $2d$ boson-fermion 
correspondence {\cite{Col,Fateev1}. 
 To obtain a well-defined QFT counterterm(s) is (are) added to cancel the 
fermion loop divergence.
Consider firstly the case $n=3$. 
For the ``minimal'' $\MMC_3$ (${m_0 \over m_1} = 1$, with constraint $(i)$ in 
$(3.2)$), after fermionization 
a massive Thirring model coupled to an $A_1^{(1)}$ ATFT is obtained.
Its dual representation is given in \cite{Fateev1}.
 Consider next the MSG model associated to $\hbox{\it I\hskip
 -2.pt M \hskip -7 pt I \hskip - 3.pt \CC}_3 \subset \hbox{\it I\hskip -2.pt M 
\hskip -7 pt I \hskip -
 3.pt \CC}_4$ (${m_0 \over m_1} = 1/2$,
case $(i)$ in (\ref{p3m})). 
For small $\beta_0=\beta$, the $2d$ fermion-boson correspondence 
\cite{Col,Fateev1} gives :
\begin{eqnarray}
{\cal{A}}^{(3|4)} &=& \int d^2x \Big[ i\overline{\psi} 
\gamma_{\mu}\partial_{\mu}\psi -\frac{g_0}{2}
(\overline{\psi}\gamma_{\mu}\psi)^2 - M\overline{\psi}\psi e^{\beta\varphi_0}
 \nonumber \\
&&\ \ \ \ \ \ \ \ \ \ \ \ + \frac{1}{2}(\partial_{\mu} \varphi_0)^2 - 
\frac{M^2}{2\beta^2} 
\big(2e^{-\beta\varphi_0}+e^{2\beta\varphi_0}\big)\Big],
\label{n=3BD}
\end{eqnarray}
with $g_0$ defined by $g_0/\pi = \frac{4 \pi}{\alpha_0^2}-1$.
In the non-minimal case ($\MMC_3 \subset \MMC_4$), 
an $A_2^{(2)}$ ATFT
(Bullough-Dodd model) is obtained for the purely bosonic part\,\footnote{Up 
to the choice
 $\mu=\frac{3}{4\beta}(2M)^
{\frac{3}{2}}$ and the shift \ $\varphi_0 \rightarrow \varphi_0-
\frac{1}{\beta}\ln(\frac{\sqrt{2M}}
{\beta})$.}.
 V. Fateev  has shown \cite{Fateev1} that 
the QFT (\ref{n=3BD}) possesses a dual representation 
associated to
the complex sinh-Gordon model (CSG) (sigma model with Witten's black hole 
metric). In this dual 
representation
 (with the complex scalar field $\chi=\chi_1+i\chi_2$), the action is
\begin{eqnarray} 
\tilde{{\cal A}}^{(3|4)} =  \int d^2x\ \left( \frac{1}{2}
\frac{\partial_{\mu}\overline{\chi}
\partial_{\mu}\chi}
{1+(\frac{\gamma}{2})^2|\chi|^2} - \frac{M^2}{2} |\chi|^2 \Big[1+
(\frac{\gamma}{2})^2|\chi|^2\Big]
\right).\label{dualn=3}
\end{eqnarray} 
It was denoted $BC_0(\chi,\gamma)$ where the coupling constant $\gamma$ is 
defined by 
$\gamma=\frac{4\pi}{\beta}$.

The minimal $\MMC_4$ model (${m_0 \over m_1}=1$, case $(i)$ in $(3.3)$)  was 
already envisaged in \cite{muli4} and its fermionization corresponds to two 
massive Thirring models coupled
to an $A_1^{(1)}$ ATFT. The dual representation is given in \cite{Fateev1}. \\

Consider next the MSG model associated to $\hbox{\it I\hskip -2.pt M \hskip -7
 pt I \hskip - 3.pt 
\CC}_5 \subset 
\hbox{\it I\hskip -2.pt M \hskip -7 pt I \hskip - 3.pt \CC}_6$ for 
$\alpha_0=\alpha_1=\alpha$ and 
$\beta_0=-\beta_1=\beta$, ${m_0 \over m_2}={m_1 \over m_2}={1 \over 2}$, 
case $(i)$ in $(3.4)$.
 For small $\beta$, the $2d$ fermion-boson correspondence gives 
($g_0=g_1=g$) \footnote{For the choice 
$\mu=\frac{5}{2}(\frac{M^{4}}{\beta^2})^{\frac{1}{3}}$ and the shifts 
$\varphi_{0} \rightarrow 
\varphi_{0} 
- \frac{1}{3\beta} \ln (\frac{M}{\beta^2})$ and $\varphi_{1} \rightarrow 
\varphi_{1} + \frac{1}{3\beta}
 \ln (\frac{M}{\beta^2})$, then the purely bosonic part possesses a stable 
classical vacuum.} :
\begin{eqnarray}
\mathcal{A}^{(5|6)} &=& \int d^2x \Big[\frac{1}{2} 
(\partial_{\mu}\varphi_{0})^2 + \frac{1}{2} 
(\partial_{\mu}
\varphi_{1})^2 + \sum_{a=0}^{1}\big[i \overline{\psi}_{a}\gamma_{\mu}
\partial^{\mu}\psi_{a}-
\frac{g}{2}(\overline
{\psi}_{a}\gamma_{\mu}\psi_{a})^2\big],\label{n=56}\\ \nonumber& &\ \ \ \ \ \ 
\ \ \  \ \ \  \  -M 
\overline{\psi}_
{0}{\psi_{0}}e^{\beta\varphi_0}  -M \overline{\psi}_{1}{\psi_{1}}
e^{-\beta\varphi_1} -\frac{M^2}{2
\beta^2}
\big(e^{2 \beta \varphi_{0}}+2e^{- \beta (\varphi_{0}-\varphi_{1})}+
e^{-2 \beta \varphi_{1}}\big)\Big],
\end{eqnarray}
which corresponds to two massive Thirring (MT) models coupled with a 
$C_2^{(1)}$ ATFT. 
It was already shown by V. Fateev \cite{Fateev1} by perturbative and 
non-perturbative 
analysis  that the QFT (\ref{n=56}) possesses a dual representation 
$\tilde{\cal A}^{(5|6)}$ \
\begin{eqnarray} 
\tilde{{\cal A}}^{(5|6)} =  \int d^2x\ \left( \sum_{a=0,1}\frac{1}{2}
\frac{\partial_{\mu}
\overline{\chi_a}\partial_{\mu}\chi_a}{1+(\frac{\gamma}{2})^2|\chi_a|^2} - 
\frac{M^2}{2}\Big[ \chi_0 
\overline{\chi_0} + \chi_1\overline{\chi_1} + (\frac{\gamma}{2})^2(\chi_0
\overline{\chi_0})(\chi_1
\overline{\chi_1})\Big]\right),\label{dualn=5,6}
\end{eqnarray} 
denoted $D_0^{(2)}(\chi_{0,1},\gamma)$ with $\gamma$ defined as above.\\
From the above results and those obtained in \cite{muli4}, the 
correspondence between known QFTs and MSG theories may be summarized in the 
following table where,
for each integrable MSG,  
its QFT in terms of MT coupled with specific ATFT and its dual QFT is 
indicated. 
In this compilation we only quote MSG models with a known dual representation 
after fernionization. 
The important fact is that all these models are generated by  MC algebras.\\ 

%
\begin{tabular}{|l|l|l r|l|}
\hline
& & & & \\
$n$ & Multicomplex & \ \ \ \ \ \ \ \ \ \ \ \ QFT &  & Dual QFT \\
& \ \ Embedding & \ \ \ \ \ \ \ \ \ \ \ \ (coupling $\beta$)  &  &  
(coupling $\gamma={{4\pi}\over{\beta}}$)\\
\hline \hline
3  & $\hbox{\it I\hskip -2.pt M \hskip -7 pt I \hskip - 3.pt \CC}_3$ & $MT$ 
coupled with $A_{1}^{(1)}$ & $(3|3)$ & CSG\\
   & $\hbox{\it I\hskip -2.pt M \hskip -7 pt I \hskip - 3.pt \CC}_3 \subset 
\hbox{\it I\hskip -2.pt M \hskip -7 pt
 I \hskip - 3.pt \CC}_4$ & $MT$ coupled with $A_{2}^{(2)}$ & $(3|4)$ & 
CSG $BC_0(\chi,\gamma)$ \\
\hline
4  & $\hbox{\it I\hskip -2.pt M \hskip -7 pt I \hskip - 3.pt \CC}_4$ & 
$MT \otimes MT$ coupled with $A_{1}^{(1)}$
 & $(4|4)$ & $N=2$ supersym-\\
& & & & metric sine-Gordon,$\hdots$\\
\hline
5  & $\hbox{\it I\hskip -2.pt M \hskip -7 pt I \hskip - 3.pt \CC}_5 \subset 
\hbox{\it I\hskip -2.pt M \hskip -7 pt I \hskip - 3.pt \CC}_6$ & $MT \otimes
 MT$ coupled with $C_{2}^{(1)}$ & $(5|6)$ & CSG $D_0^{(2)}(\chi_{0,1},\gamma)$
 \\ 
\hline
\end{tabular}
\vspace{5mm}

Clearly the analysis can be carried out for any order $n$ of the multicomplex 
space, for any set of $\{m_a\}$, 
and for a specific choice of the constraints on parameters. 
It is always possible to fermionize the corresponding MSG model as before and 
to obtain its QFT interpretation 
in terms of massive Thirrings models. For example the minimal case $\MMC_5$ 
corresponds to two massive Thirring models coupled with a $D_3^{(2)}$ ATFT. 
Its dual counterpart is an open question.
Hence this compilation is not exhaustive since integrable MSG-models may exist
 with or without a dual representation. 
The benefit of
 the embedding property 
$\hbox{\it I\hskip -2.pt M \hskip -7 pt I \hskip - 3.pt
 \CC}_n \subset \hbox{\it I\hskip -2.pt M \hskip -7 pt I \hskip - 3.pt \CC}_m$
 is that integrable models reached 
from order $n$ can also be related to MC-numbers of order $m > n$.

\mysection{ Comments, perspectives and conclusions }

For $n < 6$ different integrable models were found. Their integrability is 
ensured from different constraints
among the parameters. However when $n \ge 6$ the spin $4$ current is conserved 
(to first order in the interaction term) only
for the generic solutions (case (ii) in (2.2), (iii) in (2.3) and (ii) in 
(2.4)). Similar conclusions are found in \cite{damien} using different 
techniques.
It can be checked explicitly  that
all the conditions for current conservation (to first order) of the generic 
model ($m_a$ not fixed)  
related to the ``embedding''
$\hbox{\it I\hskip -2.pt M \hskip -7 pt I \hskip - 3.pt \CC}_n \subset 
 \hbox{\it I\hskip -2.pt M \hskip -7 pt I \hskip - 3.pt \CC}_m$ 
studied so far are given by a single formula  ($ m_a \ne 0$)
\beqa
\label{constrmn}
\left\{
\begin{array}{ll}
\sum \limits_{a=0}^{n/2-2} {m_a \over m_{{n \over 2}-1}} 
(\alpha_a^2-\beta_a^2) 
+\alpha_{n/2-1}^2- \sum \limits_{a=0}^{n/2-2}
({m_a \over m_{{n \over 2}-1}} \beta_a)^2 - \Big[ 1 + \sum 
\limits_{a=0}^{n/2-2} {m_a \over m_{{n \over 2}-1}} \Big] = 0,
&\!\!\!\!\! n\mathrm{~even} \\
& \\
\sum \limits_{a=0}^{(n-1)/2-1} {2 m_a \over m_{{n-1 \over 2}}} (\alpha_a^2-
\beta_a^2) 
- \sum \limits_{a=0}^{(n-1)/2-1}
({2 m_a \over m_{{n-1 \over 2}}} \beta_a)^2 - \Big[ 1 + \sum 
\limits_{a=0}^{(n-1)/2-1} {2 m_a \over m_{{n-1  \over 2}}} \Big] = 0,
&\!\!\! n\mathrm{~odd}. 
\end{array}
\right.
\eeqa 

 When the parameters are real and after a proper rescaling of the  
$\alpha_a$'s and $\beta_a$'s the geometric interpretation is simply that 
current conservation at least 
to the first order
occurs
if the parameters live on hyperboloids. They are invariant under a
pseudo-rotational group \cite{galice}.\\

 When imaginary parameters are allowed some MSG-models  possess
interesting dual properties in terms of coupling. Using the flexibility
of the non-minimal and non-embedded representation of the multisine it is 
possible to obtain a
 dual counterpart of a given MSG. 
Consider for instance the MSG-model associated with $n=4$. 
Any choice (integer or not) for $m_0$ 
and $m_1$ is possible when relaxing the embedding interpretation. For the set 

\begin{eqnarray}
\begin{array}{ll}
\left\{
\begin{array}{l}
\alpha_0^2-\beta_0^2= {1 \over 2} \\
\alpha_1= { i \over 2 \beta_0}
\end{array}
\right. 
&{m_0 \over m_1} = {1 \over 2 \beta_0^2}
\end{array}
\end{eqnarray}
the spin  $4$ current
\begin{eqnarray}
\label{t4m}
T^{(4|m)}&=& {3 + 10 \beta_0^2 \over 12 \beta_0^2 }\ \  (\partial \phi_0)^4 + 
             {\beta_0^2 \over 3 + 6 \beta_0^2} \ \  (\partial \phi_1)^4 + 
              {\beta_0^2 \over 3 + 6 \beta_0^2} \ \  (\partial \varphi_0)^4 
 \nonumber \\
     &+&    \Big[ (\partial \phi_0)^2 (\partial \varphi_0)^2 +
                  (\partial \phi_1)^2 (\partial \varphi_0)^2 +
                  (\partial \phi_0)^2 (\partial \phi_1)^2 \Big]   \nonumber \\
      &+&   {2 + 4 \beta_0^2 \over \beta_0} \ \  (\partial \phi_0)^2 
(\partial^2 \varphi_0) 
            -{ 1 \over \beta_0} \ \    (\partial \phi_1)^2 (\partial^2 
\varphi_0)  \\
&+& 
     { 7 + 24 \beta_0^2 + 16 \beta_0^4 \over 3 + 6 \beta_0^2}\ \ (\partial^2 
\varphi_0)^2 
      +    {3+22 \beta_0^2+56 \beta_0^4 + 32 \beta_0^6 \over 6 \beta_0^2 + 12 
\beta_0^4}\ \  (\partial^2 \phi_0)^2 \nonumber \\
   &+&      {3 + 19 \beta_0^2 + 36 \beta_0^4 + 16 \beta_0^6 \over 3 \beta_0^2 
+ 6 \beta_0^4}\ \  (\partial^2 \phi_1)^2. \nonumber
\end{eqnarray}
is  conserved. For the set
\begin{eqnarray}
\label{n=4dual}
\begin{array}{ll}
\left\{
\begin{array}{l}
\alpha_0= -i \beta_0 \\
\alpha_1^2 - {1 \over {4 \beta_0^2}}= { 1 \over 2 }
\end{array}
\right. 
&{m_0 \over m_1} = {1 \over 2 \beta_0^2}
\end{array}
\end{eqnarray}
a spin $4$ current is also conserved. It is given by Eq.($4.3$) with the 
duality transformation
$\sqrt{2} \beta_0 \longrightarrow - {1 \over{\sqrt{2} \beta_0}}$ together
 with the change  $\phi_0 \leftrightarrow \phi_1$. This model is integrable
and possesses a dual representation  in terms of complex sinh-Gordon
model coupled to a massive Thirring \cite{bf}. Its alternative truly bosonic
representation is obtained from the initial lagrangian with the change
(\ref{n=4dual}).
When $\sqrt{2} \beta_0 \longrightarrow - 
{1 \over{\sqrt{2} \beta_0}}$, $m_0/m_1 \longrightarrow 2 \beta_0^2$, it is
 always possible to find new parameters $m_a'$
such that  $m_0'/m_1' = 1/(2 \beta_0^2)$. A new generator $e^{\vee}$ (the 
``dual'' of $e$) of the  multicomplex space is then obtained from the initial 
one $e$ by the exchange $m_0 \leftrightarrow m_1$, which corresponds in terms 
of coupling to the exchange $\sqrt{2} \beta_0 \longrightarrow - 
{1 \over{\sqrt{2} \beta_0}}$.  
In other words, the duality transformation (\ref{n=4dual}) induces
a transformation among the $m_a$'s themselves,  that is
a duality transformation in the MC-algebras.\\

 One other legitimate question is whether or not  {\it any} ATFT potential
can be described within an MSG-model.
A multisine-Gordon model of order $n$ is specified by  
the set of $m_a$'s and the parameters $\{\alpha_a,\beta_a\}$. 
In the previous section, the parameters $\alpha_a, \beta_a$ were {\it real}.
 However when 
$\alpha_a \in \RR$ and $\beta_a \in i \RR$,
 different ATFT potentials can
 be reached depending on the sets of  $\{m_a\}$  and/or $\{\alpha_a,\beta_a\}$
. 
 Since in particular for $n=3$ the affine Lie algebra $A_2^{(1)}$ and
 $D_3^{(2)}$ can be obtained
 for imaginary parameters as $MSG_{(3|3)}(\beta \sqrt {3 \over 2}, i 
{\beta \over \sqrt 2})$
and  $MSG_{(3|3)}(\beta, i \beta)$ the first step is to consider the so-called
 totally compact
mus-functions based on a complex algebra. Then one way to proceed is through 
the identification 
of the vertex operators 
associated to Lie algebras with the corresponding operators obtained from
 the expansion of $\mathrm{mus}_0(\Phi)$\,\footnote{An alternative is to use
 the approach on non-local conserved charges \cite{damien}.}
 In the process the condition $\sum \limits_{a=0}^{n-1} n_a r_a = 0$, 
where $\{n_a\}$ are the Ka\v c's labels and  $\{{r}_{a}\}$ the roots of the 
affine Lie
 algebra, is in correspondence with the condition of unimodularity Eqs. 
(\ref{constr_m}). 
However some care is needed within this identification. On the one hand the 
metric
of the ATFT governs the identification of the induced metric of the MSG. It is
 in general non-invariant. On the
other hand the constraint among the MSG-fields depends in an essential way on 
the manner the Lie algebra condition 
$\sum \limits_{a=0}^{n-1} n_a r_a = 0$ is implemented. The final outcome is 
that {\it all} ATFT potentials can be related to an
MSG one. The detailed analysis is in perfect agreement with other results 
\cite{damien}.
  
To conclude a class of new multi-parameter models has been found with
a generic condition for current conservation to first order in the interaction
 term.
In order to ensure integrability of these new models, it remains to check that 
quantum corrections do not not spoil current conservation. Finding a general 
method to derive 
their dual counterparts is a challenging problem. 
 
A field theoretic formulation  in terms of minimal and
 non-minimal representation of  MC-algebra and multisine functions provides a 
unified
algebraic description  of various integrable $2d$ deformed Toda field theories.
 Thereby their 
studies is reduced to that of the underlying multi-paramater space. Some 
additional 
indications were thus gathered on
 the richness of the algebraic structure of MSG in the understanding of 
hidden symmetries 
which may appear in multi-parametrized quantum field theories.\\

 
\paragraph*{Acknowledgements}

V. A. Fateev and A. Neveu are gratefully acknowledged for useful discussions. 
We are also grateful to Ed. Corrigan for his final comments.  
P.B thanks  KIAS for its hospitality. This work is supported in part by 
European Union
contract ERBFMRX CT 960012

\end{document}